\documentclass[12pt,aps,prb,preprint]{revtex4}   

\usepackage{amsmath}    
\usepackage{amsfonts}  
\usepackage{amssymb}
\usepackage{graphicx}   

\usepackage[latin1]{inputenc}     
\usepackage[T1]{fontenc}

\usepackage{float}   

\draft

\begin{document}

\title{The Principle of Least Action as interpreted by Nature and by the Observer}

\author{Michel Gondran}
 \affiliation{University
Paris Dauphine, Lamsade, 75 016 Paris, France}
 \email{michel.gondran@polytechnique.org}   
 \author{Alexandre Gondran}
\affiliation{\'Ecole Nationale de l'Aviation Civile, 31000 Toulouse,
France}
 \email{alexandre.gondran@enac.fr}   

\begin{abstract}

In this paper, we show that the difficulties of interpretation of
the principle of least action concerning "final causes" or
"efficient causes" are due to the existence of two different
actions, the "Euler-Lagrange action" (or classical action) and the
"Hamilton-Jacobi action". These two actions, which are not clearly differentiated in the texbooks, are solutions to the
same Hamilton-Jacobi equation, but with very different initial
conditions: smooth conditions for the Hamilton-Jacobi action,
singular conditions for the Euler-Lagrange action. There are related by the Minplus Path Integral which is the analog in classical mechanics of the Feynmann Path Integral in quantum mechanics. Finally, we propose a clear-cut interpretation of the principle of least action: the
Hamilton-Jacobi action does not use "final causes" and seems
to be the action used by Nature; the Euler-Lagrange action uses "final causes" and is the action used by an observer to
retrospectively determine the trajectory of the particle.

\end{abstract}

\maketitle


\section{Introduction}
\label{sect:intro}

In 1744, Pierre-Louis Moreau de Maupertuis (1698-1759) introduced
the action and the principle of least action into classical
mechanics \cite{Maupertuis1744}: "\emph{Nature, in the
production of its effects, does so always by the simplest means
[...] the path it takes is the one by which the quantity of action
is the least,}" and in 1746, he states \cite{Maupertuis1746}:
"\emph{This is the principle of least action, a principle so wise
and so worthy of the supreme Being, and intrinsic to all natural
phenomena [...] When a change occurs in Nature, the quantity of
action necessary for change is the smallest possible. The quantity
of action is the product of the mass of the body times its
velocity and the distance it moves.}" Maupertuis understood that,
under certain conditions, Newton's equations imply the
minimization of a certain quantity. He dubbed this quantity the
action. Euler \cite{Euler1744}, Lagrange \cite{Lagrange},
Hamilton \cite{Hamilton1834}, Jacobi
\cite{Jacobi} and others, turned this principle of least action
into one of the most powerful tools for discovering the laws of
nature\cite{Feynman,Landau}. This principle serves to determine
the equations of themotion of a particle (if we minimize the
trajectories) and the laws of nature (if we minimize the
parameters defining the fields).

However, when applied to the study of particle trajectories, this
principle has often been viewed as puzzling by many scholars,
including Henri Poincar\'e, who was nonetheless one of its most
intensive users \cite{Poincare}: "\emph{The very enunciation of
the principle of least action is objectionable. To move from one
point to another, a material molecule, acted on by no force, but
compelled to move on a surface, will take as its path the geodesic
line, i.e., the shortest path . This molecule seems to know the
point to which we want to take it, to foresee the time it will
take to reach it by such a path, and then to know how to choose
the most convenient path. The enunciation of the principle
presents it to us, so to speak, as a living and free entity. It is
clear that it would be better to replace it by a less
objectionable  enunciation, one in which, as philosophers would
say, final effects do not seem to be substituted for acting
causes.}"

We will show that the difficulties of interpretation of the
principle of least action concerning the "final causes" or the
"efficient causes" are due to the existence of two different
actions: the "Euler-Lagrange action" (or classical action)
$S_{cl}(\mathbf{x},t;\mathbf{x}_0)$, which links the initial
position $\mathbf{x}_0$ and its position $\mathbf{x}$ at time t,
and the "Hamilton-Jacobi action" $S(\mathbf{x},t)$, which depends
on an initial action $S_{0}(\mathbf{x})$.

These two actions are solutions to the same Hamilton-Jacobi
equation, but with very different initial conditions: smooth
conditions for the Hamilton-Jacobi action, singular conditions for
the Euler-Lagrange action. These initial conditions are not taken
into account in classical mechanics textbooks \cite{Landau,Goldstein}. We show that they are the key to
understanding the principe of least action.

In section~\ref{sect:ELHJ}, we recall the main properties of the Euler-Lagrange
action and we propose a new interpretation.

In section~\ref{sect:HJ}, we propose a novel way to look at the
Hamilton-Jacobi action, in explaining the principle of least
action that it satisfies.

In section~\ref{sect:Minplus}, we show how Minplus analysis, a new branch of
nonlinear mathematics, explains the difference between the
Hamilton-Jacobi action and the Euler-Lagrange action. The equation between these two actions, which we call the Minplus
Path Integral, is the
analog in classical mechanics of the Feynman Path integral in
quantum mechanics. 

In conclusion, we respond to Poincar\'e by provinding a clear-cut
interpretation of this principle.

\section{The Euler-Lagrange  action}
\label{sect:ELHJ}

Let us consider a system evolving from the position
$\textbf{x}_{0}$ at initial time to the position $\textbf{x}$ at
time $t$ where the variable of control \textbf{u}(s) is the
velocity:
\begin{eqnarray}\label{eq:evolution}
\frac{d \textbf{x}\left( s\right) }{ds}=\mathbf{u}(s),\qquad\forall s\in\left[ 0,t\right]\\
\label{eq:condinitiales}
\textbf{x}(0) =\mathbf{x}_{0},\qquad\textbf{x}(t) =\mathbf{x}.
\end{eqnarray}

If $L(\textbf{x},\dot{\textbf{x}},t)$ is the Lagrangian of the
system, when the two positions $\textbf{x}_0$ and $\textbf{x}$ are
given, \emph{the Euler-Lagrange action} $S_{cl}(\mathbf{x},t;
\textbf{x}_0) $ is the function defined by:
\begin{equation}\label{eq:defactioncondit}
S_{cl}(\textbf{x},t;\mathbf{x}_{0})=\min_{\textbf{u}\left(
s\right),0 \leq s\leq t} \int_{0}^{t}L(\textbf{x}(s),
\textbf{u}(s),s)ds,
\end{equation}
where the minimum (or more generally the minimum or the saddle point\cite{Gray}) is taken on the
controls $\mathbf{u}(s)$, $s\in$ $\left[ 0,t\right]$, with the
state $\textbf{x}(s)$ given by equations (\ref{eq:evolution}) and
(\ref{eq:condinitiales}). This is the principle of least action
defined by Euler \cite{Euler1744} in 1744 and Lagrange \cite{Lagrange} in 1755. 

The solution $(\widetilde{\textbf{x}}(s),\widetilde{\textbf{u}}(s)
)$ of (\ref{eq:defactioncondit}), if the Lagrangian
$L(\textbf{x},\dot{\textbf{x}},t) $ is twice differentiable,
satisfies the Euler-Lagrange equations on the interval $[0,t] $:
\begin{equation}\label{eq:EulerLagrange1}
\frac{d}{ds}\frac{\partial L}{\partial
\dot{\textbf{x}}}(\textbf{x}(s),\dot{\textbf{x}}(s),s)-
\frac{\partial L}{\partial
\textbf{x}}(\textbf{x}(s),\dot{\textbf{x}}(s),s)=0,\quad\forall s\in\left[ 0,t\right]
\end{equation}
\begin{equation}\label{eq:EulerLagrange12}
\textbf{x}(0) =\mathbf{x}_{0},~~~~\textbf{x}(t) =\mathbf{x}.
\end{equation}
For a non-relativistic particle in a linear potential field with
the Lagrangian $L(\mathbf{x},\mathbf{\dot{x}},t)= \frac{1}{2}m
\mathbf{\dot{x}}^2 + \textbf{K}. \textbf{x}$, equation
(\ref{eq:EulerLagrange1}) yields $\frac{d}{ds}( m
\dot{\textbf{x}}(s)) - \textbf{K}=0 $. We successively obtain
$\dot{\widetilde{\textbf{x}}}(s)= \widetilde{\textbf{v}}_0 + \frac{\textbf{K}}{ m}s$, 
$\widetilde{\textbf{x}}(s)= \textbf{x}_0 +
\widetilde{\textbf{v}}_0 s + \frac{\textbf{K}}{2 m} s^{2}$. The initial velocity $\widetilde{\textbf{v}}_0$ is defined using $\widetilde{\textbf{x}}(t) =\mathbf{x} $ (equation (\ref{eq:EulerLagrange12})) in the last equation. We obtain $\widetilde{ \textbf{v}}_0=\dfrac{\textbf{x}-\textbf{x}_0}{2t}-\dfrac{K}{2m}t$.
The integration constant $\widetilde{\textbf{v}}_0$ can be defined only if we know the position $\textbf{x}$ of the particle at time t.
Finally, the trajectory minimizing
the action is $\widetilde{\textbf{x}}(s)= \textbf{x}_0 +
\frac{s}{t} (\textbf{x} -\textbf{x}_0)- \frac{\textbf{K}}{2 m} t s
+ \frac{\textbf{K}}{2 m} s^2$ and the Euler-Lagrange action is
equal to
\begin{equation}\label{eq:actionlinearEulerLagrange}
S_{cl}( \mathbf{x},t; \textbf{x}_0)= m
\frac{(\textbf{x}-\textbf{x}_0)^2}{2 t}+ \frac{K .(\textbf{x} +
\textbf{x}_0)}{2}t - \frac{K^2}{24 m}t^3.
\end{equation}

\begin{figure}[h]
\includegraphics[width=0.6\linewidth]{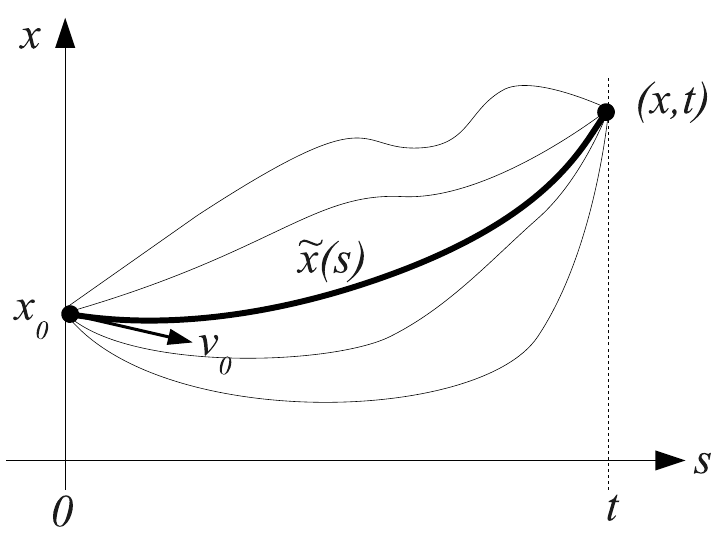}
\caption{\label{fig:trajEL} Different trajectories $\textbf{x}(s) $ ($0\leq
s \leq t$) between $(\textbf{x}_0,0)$ and $(\textbf{x},t)$ and the optimal trajectory $\widetilde{\textbf{x}}(s)$ 
with $\widetilde{\textbf{v}}_0= \frac{\textbf{x}-\textbf{x}_0}{t}-\frac{K t}{2m}$.}
\end{figure}

Figure \ref{fig:trajEL} shows different trajectories going from $\textbf{x}_0$ at time 
$t=0$ to $\textbf{x}$ at final time $t$. The parabolic trajectory $\widetilde{\textbf{x}}(s) $ 
corresponds  to the trajectory that realizes the minimum in equation (\ref{eq:defactioncondit}).

Equation (\ref{eq:defactioncondit}) seems to show that, among the
trajectories that can reach ($\textbf{x},t$) from the initial
position $\textbf{x}_0 $, the principle of least action allows to
choose the velocity at each time. In reality, the principle of
least action used in equation (\ref{eq:defactioncondit}) does not
choose the velocity at each time $s$ between $0$ and $t$, but only when
the particle arrives at $\textbf{x}$ at time $t$. The knowledge
of the velocity at each time $s$ ($0\leq
s \leq t$) requires the resolution of the
Euler-Lagrange equations
(\ref{eq:EulerLagrange1},\ref{eq:EulerLagrange12}) on the whole
trajectory. In the case of a non-relativistic particle
in a linear potential field, the velocity at time $s$ ($0\leq
s \leq t$) is $\widetilde{\textbf{v}}(s)= \frac{\textbf{x}-\textbf{x}_0}{ t} - \frac{K t}{2 m}+\frac{K s}{ m}$ with the initial velocity
\begin{equation}\label{eq:vitesseEulerinit}
\widetilde{\textbf{v}}_0= \frac{\textbf{x}-\textbf{x}_0}{ t}- \frac{K t}{2 m}.
\end{equation}
Then, $\widetilde{\textbf{v}}_0$ depends on the position $\textbf{x}$ of the particle at the final
time $t$. This dependence of the "final causes" is
general. This is Poincar\'e's main criticism of the principle of
least action: "\emph{This molecule seems to know the point to
which we want to take it, to foresee the time it will take to
reach it by such a path, and then to know how to choose the most
convenient path.}"

One must conclude that, without knowing the initial velocity, the
Euler-Lagrange action answers a problem posed by an observer, and
not by Nature: "What would be the velocity of the particle at the
initial time to attained $\textbf{x}$ at time $t$?" The resolution
of this problem implies that the observer solves the
Euler-Lagrange equations
(\ref{eq:EulerLagrange1},\ref{eq:EulerLagrange12}) after the
observation of $\textbf{x}$ at time $t$. This is an \emph{a
posteriori} point of view.

\section{The hamilton-jacobi action}
\label{sect:HJ}

The Hamiton-Jacobi action will overcome this \textit{a priori} lack of  knowledge of the initial velocity in the Euler-Lagrange action. Indeed, at the initial time, the Hamilton-Jacobi action $S_0(\textbf{x})$ is known. The knowledge of this initial action $S_0(\textbf{x})$ involves the knowledge of the velocity field at the initial time that satisfies $\textbf{v}_0(\textbf{\textbf{x}})=\dfrac{\nabla S_0(\textbf{x})}{m}$. \textit{The Hamilton-Jacobi action } $S(\mathbf{x},t) $ at $\textbf{x} $ and time t is then the function defined by:
\begin{equation}\label{eq:defactionHJ}
S(\mathbf{x},t)=\min_{\textbf{x}_0;\mathbf{u}\left( s\right),0
\leq s\leq t }\left\{ S_{0}\left( \mathbf{x}_{0}\right)
+\int_{0}^{t}L(\textbf{x}(s), \mathbf{u}(s),s)ds\right\}
\end{equation}
where the minimum is taken on all initial positions $\textbf{x}_0$
and on the controls $\mathbf{u}(s)$, $s\in$ $\left[ 0,t\right]$,
with the state $\textbf{x}(s)$ given by the equations
(\ref{eq:evolution})(\ref{eq:condinitiales}).

This Hamilton-Jacobi action with its initial solution $S_0(\textbf{x})$ is known in the mathematical texbooks\cite{Lions, Evans} for optimal control problems, but is ignored in the physical textbooks such as those of Landau\cite{Landau} chap.7 § 47 and Goldstein\cite{Goldstein} chap.10 where there is no mention of the initial condition $S_0(\textbf{x})$. It is often confused in the texbooks with what is refered to as the principal function of Hamilton.

The initial condition $S_{0}(\textbf{x}) $ is 
mathematically necessary to obtain the general solution to the
Hamilton-Jacobi equations (\ref{eq:HJ})(\ref{eq:condinitialHJ}). Physically, it is the condition that describes the preparation of the particles. We will see that this
initial condition is the key to understanding the least action principle.

Noting that because $S_0(\textbf{x}_0)$ does not play a role in
(\ref{eq:defactionHJ}) for the minimization on $\textbf{u}(s)$, we
obtain a relation between the Hamilton-Jacobi action and
Euler-Lagrange action:
\begin{equation}\label{eq:ELHJ}
S(\mathbf{x},t)=\min_{\textbf{x}_0} ( S_{0}\left(
\mathbf{x}_{0}\right) + S_{cl}(\textbf{x},t;\textbf{x}_0) ).
\end{equation}
It is an equation that generalizes the Hopf-Lax and Lax-Oleinik formula,
\cite{Lions, Evans} $S(\mathbf{x},t)=\min_{\textbf{x}_0} ( S_{0}\left(
\mathbf{x}_{0}\right) + m \dfrac{(x-x_0)^2}{2t})$ that corresponds to the particular case of the free particle where the Euler-Lagrange action is equal to $ m \dfrac{(x-x_0)^2}{2t}$.

For a particle in a linear potential $V(\textbf{x})= - \textbf{K}
.\textbf{x}$ with the initial action $S_0(\textbf{x})= m
\textbf{v}_0 \cdot \textbf{x}$, we deduce from the equation
(\ref{eq:ELHJ}) that the Hamilton-Jacobi action is equal to $
S\left( \mathbf{x},t\right)=m \textbf{v}_0 \cdot \textbf{x} -
\frac{1}{2} m \textbf{v}_0^2 t +\textbf{K}.\textbf{x} t -
\frac{1}{2} \textbf{K}.\textbf{v}_{0} t^{2} - \frac{\textbf{K}^2
t^3}{6 m}$.

\begin{figure}[h]
\includegraphics[width=0.6\linewidth]{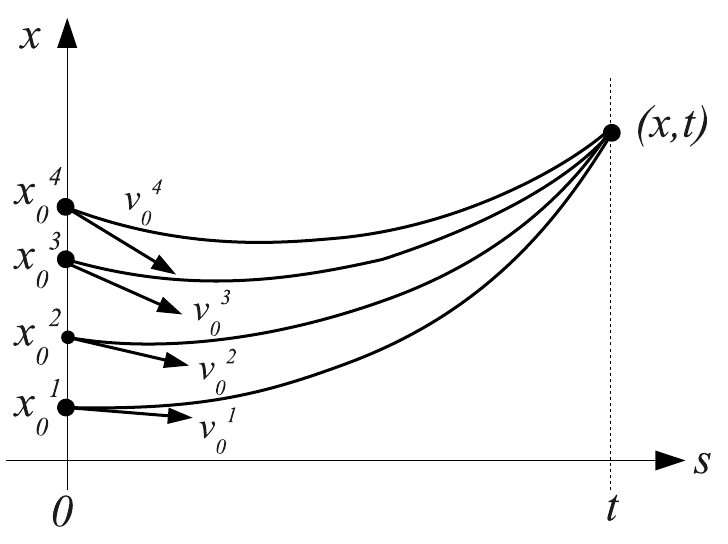}
\caption{\label{fig:trajHJ} Classical trajectories $\widetilde{x}(s) $ ($0\leq
s \leq t$) between the different initial positions $x^i_0$ and the position $x$ at time $t$. 
We obtain $\widetilde{\textbf{v}}^i_0= \frac{\textbf{x}-\textbf{x}^i_0}{ t}- \frac{K t}{2 m}$.}
\end{figure}

Figure \ref{fig:trajHJ} shows the classical trajectories (parabols) 
going from different starting points $x^i_0$ at time $t=0$ to the point $x$ at final time $t$. 
The Hamilton-Jacobi action is compute with these trajectories in equation (\ref{eq:ELHJ}).

We are now in position to derive the Hamilton-Jacobi equations.
The action $S(\mathbf{x},t)$ defined by (\ref{eq:defactionHJ}) can
be decomposed into
\begin{equation*}
S(\mathbf{x},t)=\underset{\textbf{x}_0;\mathbf{u}\left( s\right),0
\leq s\leq t }{\min
}\left\{ S_{0}\left( \mathbf{x}_{0}\right) +\int_{0}^{t -dt}L(\textbf{x}(s),%
\mathbf{u}(s),s)ds +\int_{t - dt}^{t}L(\textbf{x}(s),%
\mathbf{u}(s),s)ds\right\}
\end{equation*}
and then satisfies,between the time t-dt and t, the optimality equation:
\begin{equation*}
S(\mathbf{x},t)=\underset{\mathbf{u}\left(s\right) , t-dt\leq
s\leq t}{\min }\left\{ S(\mathbf{x}- \int^{t}_{t-dt}\textbf{u}(s)
ds,t-dt )
+\int_{t-dt}^{t}L(\textbf{x}(s),\mathbf{u}(s),s)ds\right\}.
\end{equation*}

Assuming that S is differentiable in \textbf{x} and t, L
differentiable in \textbf{x}, \textbf{u} and t, and \textbf{u}(s)
continuous, this equation becomes:
\begin{equation}\label{eq:defactionlocale}
S(\mathbf{x},t)=\underset{\mathbf{u}\left(t\right)}{\min }\left\{ S(\mathbf{x}-\textbf{u}(t)
dt,t- dt) + L(\textbf{x},
\mathbf{u}(t),t)dt +\circ \left( dt\right) \right\}.
\end{equation}
\begin{equation*}
0=\underset{\mathbf{u}\left( t\right) }{\min }\left\{
-\nabla S \left( \mathbf{x} \mathbf{,}t\right)
\textbf{u}\left( t\right) dt-\frac{\partial S}{\partial \mathbf{t%
}}\left( \mathbf{x} \mathbf{,}t\right) dt+L(\mathbf{x},
\mathbf{u}(t),t)dt+\circ \left( dt\right) \right\}
\end{equation*}
and dividing by $dt$,
\begin{equation*}
\frac{\partial S}{\partial \mathbf{t}}\left( \mathbf{x,}t\right)
=\underset{ \mathbf{u}}{\min }\left\{
L(\mathbf{x},\mathbf{u},t)-\mathbf{u\cdot }
\nabla S \left( \mathbf{x,}t\right)
\right\}.
\end{equation*}

For a convex function $f(\mathbf{u}):%
\mathbf{u}\in \mathbb{R}^{n}\rightarrow \mathbb{R}$, we recall that
we can associate its
\textit{Fenchel-Legendre transform} $\ \widehat{f}(\mathbf{r}):\mathbf{r}\in \mathbb{%
R}^{n}\rightarrow \mathbb{R}$ defined by $
\widehat{f}(\mathbf{r})=\underset{\mathbf{u\in }\mathbb{R}^{n}}{\max }(%
\mathbf{r}\cdot \mathbf{u}-f(\mathbf{u}))$.~\cite{Evans} The
Hamiltonian $H(\mathbf{x},\mathbf{p},t)$\textit{\ } is then the
Fenchel-Legendre transform of the Lagrangian
$L(\mathbf{x},\mathbf{u},t)$ for the variable \textit{\
}$\mathbf{u}$.

\emph{Then, the Hamilton-Jacobi action satisfies the
Hamilton-Jacobi equations:}
\begin{equation}\label{eq:eqgenHJ1}
\frac{\partial S}{\partial t}+H(\mathbf{x},\nabla
S {\LARGE ,}t)=0
\end{equation}
\begin{equation}\label{eq:eqgenHJ2}
S\left( \mathbf{x},0\right) =S_{0}\left( \mathbf{x}\right).
\end{equation}

For the Lagrangian $L(\mathbf{x},\mathbf{\dot{x}},t)= \frac{1}{2}m
\mathbf{\dot{x}}^2 - V(\textbf{x},t)$, we deduce the well-known
result\cite{Evans}:

\emph{The velocity of a non-relativistic classical particle  is given for each point} $ \left(
\mathbf{x,}t\right)$ \emph{by}:
\begin{equation}\label{eq:eqvitesse}
\mathbf{v}\left( \mathbf{x,}t\right) =\frac{\mathbf{\nabla }S\left( \mathbf{%
x,}t\right) }{m}
\end{equation}
\textit{where} $S\left( \mathbf{x,}t\right) $\textit{\ is the
Hamilton-Jacobi action, solution to the Hamilton-Jacobi
equations:}
\begin{equation}\label{eq:HJ}
\frac{\partial S}{\partial t}+\frac{1}{2m}(\nabla
S)^{2}+V(\textbf{x},t)=0
\end{equation}
\begin{equation}\label{eq:condinitialHJ}
S(\textbf{x},0)=S_{0}(\textbf{x}).
\end{equation}

Equation (\ref{eq:eqvitesse}) shows that the solution $S\left(
\mathbf{x,}t\right) $ to the Hamilton-Jacobi equations yields the
velocity field  for each point ($\textbf{x},t$) from the velocity
field $\frac{\nabla S_0(\textbf{x})}{m} $ at initial time. In
particular, if at initial time, we know the initial position
$\textbf{x}_{init}$ of a particle, its velocity at this time is
equal to $\frac{\nabla S_0(\textbf{x}_{init})}{m}$. From the
solution $ S\left( \mathbf{x,}t\right)$ of the Hamilton-Jacobi
equations, we deduce with (\ref{eq:eqvitesse}) the trajectories of
the particle. The Hamilton-Jacobi action $S\left(
\mathbf{x,}t\right)$ is then a field that "pilots" the particle.

\begin{figure}[h]
\includegraphics[width=0.7\linewidth]{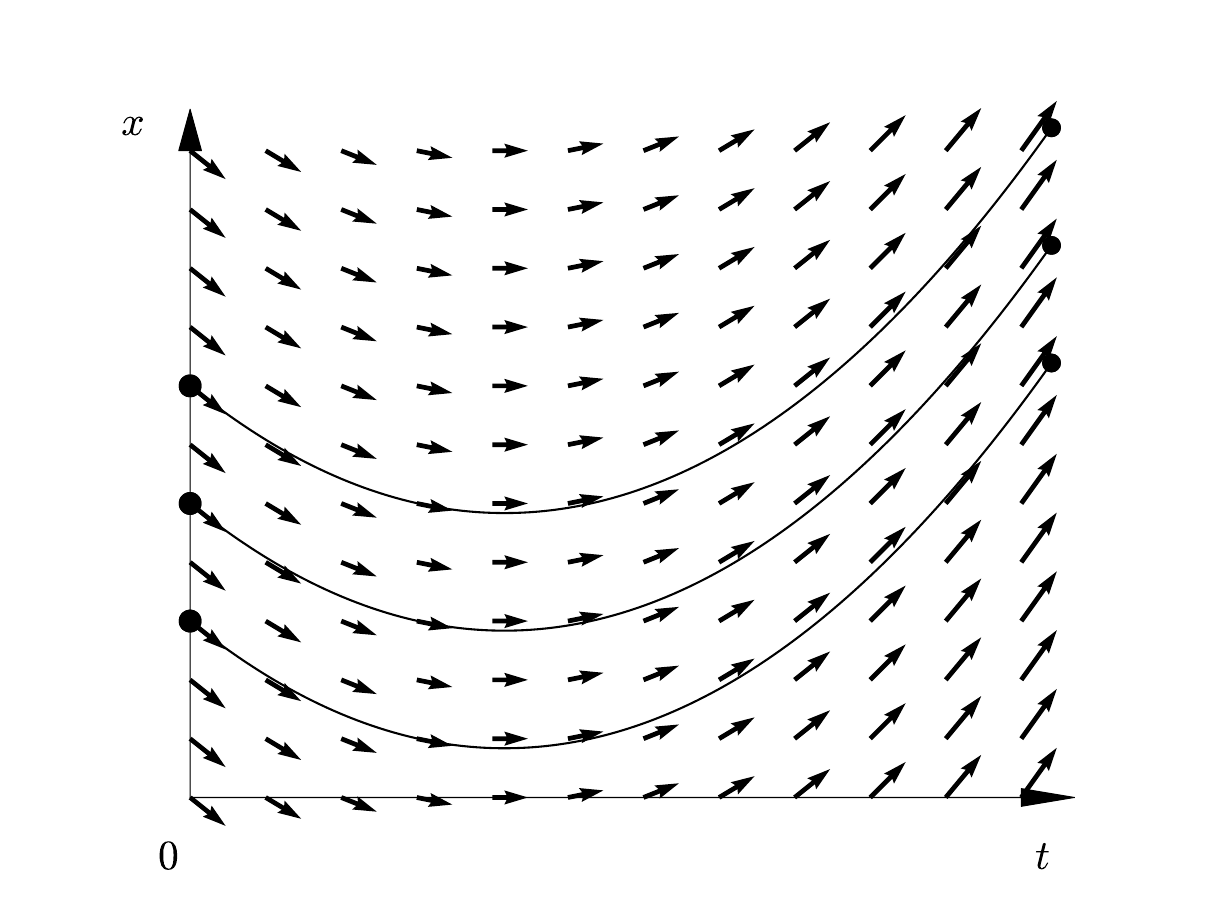}
\caption{\label{fig:fieldHJ} Velocity field that corresponds to the Hamilton-Jacobi action $ S\left( \mathbf{x},t\right)=m \textbf{v}_0 \cdot
\textbf{x} - \frac{1}{2} m \textbf{v}_0^2 t +\textbf{K}.\textbf{x}
t - \frac{1}{2} \textbf{K}.\textbf{v}_{0} t^{2} -
\frac{\textbf{K}^2 t^3}{6 m}$ ($\textbf{v}(\textbf{x},t)= \frac{\mathbf{\nabla }S\left( \mathbf{x},t\right)}{m}= \textbf{v}_0+ \frac{\textbf{K} t}{m}$) and three trajectories of particles piloted by this field.}
\end{figure}

For a particle in a linear potential $V(\textbf{x})= - \textbf{K}
.\textbf{x}$ with the initial action $S_0(\textbf{x})= m
\textbf{v}_0 \cdot \textbf{x}$, the initial velocity field is constant, 
$\textbf{v}(\textbf{x},0)= \frac{\mathbf{\nabla }S_{0}\left( \mathbf{x}\right)}{m}= \textbf{v}_0$ 
and the velocity field at time $t$ is also constant, 
$\textbf{v}(\textbf{x},t)= \frac{\mathbf{\nabla }S\left( \mathbf{x},t\right)}{m}= \textbf{v}_0+ \frac{\textbf{K} t}{m}$. 
Figure \ref{fig:fieldHJ} shows these velocity fields.


Equation (\ref{eq:eqvitesse}) seems to show that, among the
trajectories that can reach ($\textbf{x},t$) from an unknown
initial position and a known initial velocity field, Nature
chooses the initial position and at each time the velocity that
yields the minimum (or the extremum) of the Hamilton-Jacobi
action.

Equations (\ref{eq:eqvitesse}), (\ref{eq:HJ}) and
(\ref{eq:condinitialHJ}) confirm this interpretation. They show
that the Hamilton-Jacobi action $S(\mathbf{x},t)$ does not only solve a given problem with a single initial condition $\left(
\mathbf{x}_{0}, \frac{\mathbf{\nabla }S_{0}\left(
\mathbf{x}_{0}\right) }{m}\right) $, but a set of problems with an
infinity of initial
conditions, all the pairs $\left( \mathbf{y},%
\frac{\mathbf{\nabla }S_{0}\left( \mathbf{y}\right) }{m}\right)$.
It answers the following question: "If we know the action (or the
velocity field) at the initial time, can we determine the action
(or the velocity field) at each later time?" This problem is
solved sequentially by the (local) evolution equation
(\ref{eq:HJ}). This is an \emph{a priori} point of view. It is the
problem solved by Nature with the principle of least action.

Mathematical analysis can help us now to explain the differences between Hamilton-Jacobi and Euler-Lagrange actions.

\section{ Minplus analysis and the Minplus Path Integral}
\label{sect:Minplus}

There exists a new branch of mathematics, Minplus analysis,
which studies nonlinear problems through a linear approach, cf.
Maslov~\cite{Maslov,Maslov2} and
Gondran.~\cite{Gondran_1996,GondranMinoux} The idea is to
substitute the usual scalar product $\int_{X} f(x) g(x) dx$ by the
Minplus scalar product:
\begin{equation}
    (f,g) =\,\underset{x\in X}{\inf }\left\{ f(x)+g(x) \right\}
\end{equation}
In the scalar product we replace the field of the real number $(
\mathbb{R},+,\times )$ with the algebraic structure
\textit{Minplus} $( \mathbb{R}\cup \{ +\,\infty \} ,\min ,+)$,
i.e. the set of real numbers (with the element infinity $\{
+\infty \}$) equipped with the operation Min (minimum of two
reals), which replaces the usual addition, and with the operation
+ (sum of two reals), which replaces the usual multiplication.
The element $\{+\,\infty \}$ corresponds to the neutral element
for the operation Min, Min$( \{ +\infty \} ,a) =a$ $\forall a\in
\mathbb{R}$.

This approach bears a close similarity to \emph{the theory of
distributions for the nonlinear case}; here, the operator is
"linear" and continuous with respect to the Minplus structure,
though \emph{nonlinear} with respect to the classical structure $%
\left( \mathbb{R},+,\times \right)$. In this Minplus structure,
the Hamilton-Jacobi equation is linear, because if
$S_1(\textbf{x},t)$ and $S_2(\textbf{x},t)$ are solutions to
(\ref{eq:HJ}), then $\min\{\lambda + S_1(\textbf{x},t), \mu +
S_2(\textbf{x},t)\}$ is also a solution to the Hamilton-Jacobi
equation (\ref{eq:HJ}).

The analog to the Dirac distribution $\delta(\textbf{x})$ in
Minplus analysis is the nonlinear distribution
$\delta_{\min}(\textbf{x})=\{ 0~if~\textbf{x}=\textbf{0}, +\infty~
if~not\}$. With this nonlinear Dirac distribution, we can define
elementary solutions as in classical distribution theory. In
particular, we have:

\emph{The classical Euler-Lagrange action
$S_{cl}(\textbf{x},t;\textbf{x}_0)$ is the elementary solution to
the Hamilton-Jacobi equations
(\ref{eq:HJ})(\ref{eq:condinitialHJ}) in the Minplus analysis with
the initial condition}

\begin{equation}\label{eq:conditinitEL}
S_0(\mathbf{x})= \delta_{\min}(\textbf{x}- \textbf{x}_0)= \left\{\begin{array}{ll}
                  0&\textmd{if}\quad \mathbf{x}=\mathbf{x}_0,\\
		  +\infty&\textmd{otherwise}
                 \end{array}
\right.
\end{equation}

The Hamilton-Jacobi action $S(\textbf{x},t)$ is then given by the
Minplus integral
\begin{equation}\label{eq:valactionglobale}
S(\textbf{x},t)=\inf_{\textbf{x}_0} \{ S_0(\textbf{x}_0)
 + S_{cl}(\textbf{x},t;\textbf{x}_0)\}.
\end{equation}
that we call the \textit{Minplus Path Integral}. This equation is 
in analogy with the solution to the heat transfer equation given
by the classical integral:
\begin{equation}
        S(x,t)=\int S_{0}( x_0) \frac{1}{2\sqrt{\pi t}} e^{-\frac{
        \left( x-x_0\right) ^{2}}{4t}}dx_0,
\end{equation}
which is the product of convolution of the initial condition
$S_{0}(x) $ with the elementary solution to the heat transfer
equation $ e^{-\frac{x^{2}}{4t}}$.

This Minplus Path Integral yields a very simple relation between
the Hamilton-Jacobi action, the general solution to the
Hamilton-Jacobi equation, and the Euler-Lagrange actions, the
elementary solutions to the Hamilton-Jacobi equation. We can also consider that the Minplus integral
(\ref{eq:valactionglobale}) for the action in classical mechanics
is analogous to the Feynmann path integral for the wave function
in quantum mechanics. Indeed, in the Feynman path integral \cite{Feynman} (p. 58),
the wave function $\Psi(\textbf{x},t)$ at time $t$ is written as a
function of the initial wave function $\Psi_{0}(\textbf{x})$:
\begin{equation}\label{eq:interFeynman}
\Psi(\textbf{x},t)= \int F(t,\hbar)
\exp\left(\frac{i}{\hbar}S_{cl}(\textbf{x},t;\textbf{x}_{0}\right)
\Psi_{0}(\textbf{x}_{0})d\textbf{x}_0
\end{equation}
where $F(t,\hbar)$ is an independent function of $\textbf{x}$ and
of $\textbf{x}_{0}$.

The Minplus analysis have
many applications in physics: it sets the
correspondence between the Lagrangian and the Hamiltonian of a
physical system; it sets the correspondence between microscopic
and macroscopic models \cite{GondranMinoux}; it is also at the basis of Minplus-wavelets to compute Hölder exponents for fractal and multifractal functions.\cite{GondranKenoufi, GondranKenoufi2}

\section{Conclusion}
\label{sect:conclusion}

The introduction of the Hamilton-Jacobi action highlights the
importance of the initial action $ S_{0}(\mathbf{x})$, while
textbooks do not clearly differentiate these two actions. 

We are now in a position to solve Poincar\'e's puzzle: the
Hamilton-Jacobi action does not use "final causes" and seems
to be the action used by Nature; the Euler-Lagrange action uses "final causes" and is the action used by an observer to
retrospectively determine the trajectory of the particle and its
initial velocity. It is as if each time Nature solved the
Hamilton-Jacobi equation.

The principle of least action used by the Nature is represented by
equation (\ref{eq:defactionHJ}); the principle of least action
used by the observer is represented by equation
(\ref{eq:defactioncondit}). Equation 
(\ref{eq:ELHJ}) is the Minplus Path Integral that relates these two actions.

Finally, this interpretation of Euler-Lagrange action shows that the observer exists not only in quantum mechanics, but also in classical one.

\end{document}